\newcommand{\intk}{\int\!\!\!\!\!\int\!\!d^2{\bf k}_{\perp}}

\documentclass{PoS}
\usepackage{amsmath,amssymb}

\title{Integrated and unintegrated PDFs  and GPDs from effective two-body light-cone wave functions}

\ShortTitle{Exclusive hard reactions and QCD}

\author{
\speaker{Dieter M\"uller}\thanks{This work is
partly supported by the  BMBF grant no.~06BO9012, and
the Joint Research Activity ``Study of Strongly Interacting
Matter'' (acronym HadronPhysics3, Grant Agreement
no.~283286) under the Seventh Framework Programme
of the European Community.}
\\
Institute of Theoretical Physics II, Ruhr-University Bochum, 44780 Bochum, Germany\\
E-mail: \email{dieter.mueller@tp2.rub.de}
}

\author{Dae~Sung~Hwang
\thanks{This work is partly supported by
Korea Foundation for International Cooperation of Science \& Technology (KICOS)
and Basic Science Research Program through the National Research Foundation of Korea (2012-0002959).}
\\
Department of Physics, Sejong University, Seoul 143--747, South Korea \\
E-mail: \email{dshwang@sejong.ac.kr}
}

\abstract{
We suggest a classification scheme for  parton distribution models, clarify the geometrical
origin of unintegrated parton distribution  relations, which were observed in various models, present new model relations, and
provide for a so-called ``spherical'' model the analogous constraints for generalized parton distributions. Our findings suggest
that various classes of uPDF and GPD models can be obtained from effective two-body light-cone wave functions.
}

\FullConference{6th International Conference on Quarks and Nuclear Physics,  QNP 2012,\\
		April 16-20, 2012, Palaiseau, France}

\begin{document}

\section{Introduction}

\noindent
A popular and most programmatic framework, which may allow to get some insights into non-perturbative aspects of Quantum Chromodynamics (QCD), is to employ quark models, sometimes dressed with  a gauge link. Such models are utilized  to evaluate non-perturbative quantities such as form factors, parton distribution functions (PDFs), and generalized parton distributions (GPDs).
Thereby, one often assumes that, e.g., the proton at a low resolution scale can be described by naive constituent quark models and that perturbative evolution may be applied in the non-perturbative region to dynamically obtain  parton distributions that can be employed in the perturbative factorization framework. This idea, arising in the early stage of QCD, has been adopted for PDF parameterizations \cite{GluRey77} and it is safe to state that in its pure form it failed  \cite{GluReyVog98}. In order to reproduce phenomenologically acceptable results, we employ quark models as tools that are not necessarily connected with a low resolution scale. We formulate such models in terms of ``effective'' two-body light-cone wave functions (LCWFs) and parameterize them in a most flexible manner so that they can be employed in a global fitting procedure to experimental measurements of inclusive and exclusive hadronic processes. To do so one needs a building set for LCWFs
that respect the underlying Poincar{\'e} invariance of the theory, which allows us to model GPDs in terms of partonic number conserved LCWF overlaps \cite{HwaMue07}. Of course, once one has some model framework at hand one can also consistently evaluate two-quark correlation functions, e.g., so-called Wigner distributions \cite{BelJiFen03}, that are not accessible by experiments and are at present mostly discussed for forward kinematics (at least it is not shown that they provide GPDs that respect Poincar{\'e} invariance). Note that we distinguish between transverse momentum dependent parton distributions (TMDs), which absorb non-perturbative soft factors \cite{Col11}, and unintegrated PDFs (uPDFs), considered here, that have a pure operator definition.
Note that PDF evolution arises from transverse momentum integration.

There are various frameworks to set up quark models and then the following question arises:  Should one consider certain model results as being equivalent? Indeed, it was realized that in various models linear and quadratic relations among uPDFs appear, see mini review \cite{Avaetal09}. This observation was explained by rotation symmetry \cite{LorPas11}. In the following we directly utilize the spin density matrices for both uPDFs and GPDs, and their LCWFs overlap representations to set up  classification schemes that emphasize the geometrical nature of these relations.

\noindent
\section{Classification scheme of quark models}

\subsection{uPDF models}

\noindent
The  uPDFs that appear to leading power in the description of semi-inclusive deep inelastic scattering can be put into a hermitian
$4\times4$ semi-positive definite spin density  matrix with trace $2 f_1$:
\begin{eqnarray}
\label{tPhi}
 \mbox{\boldmath $\widetilde \Phi$}(x,{\bf k}_\perp) = \!\left(\!\!\!\!
\begin{array}{cccc}
\frac{f_1+g_1}{2} & \frac{|{\bf k}_\perp|e^{i\varphi}}{M}\, \frac{h_{1{\rm L}}^\perp-i h_{1}^\perp}{2} &
\frac{|{\bf k}_\perp| e^{-i\varphi}}{M}\,\frac{g^\perp_{1{\rm T}}+i f_{1{\rm T}}^\perp}{2} & h_1
\\
\frac{|{\bf k}_\perp|e^{-i\varphi}}{M}\, \frac{h_{1{\rm L}}^\perp+i h_{1}^\perp}{2} &  \frac{f_1-g_1}{2} &
 \frac{{\bf k}_\perp^2e^{-i2\varphi}}{2M^2}   h_{1{\rm T}}^\perp &
\frac{-|{\bf k}_\perp|e^{-i\varphi}}{M}\, \frac{g^\perp_{1{\rm T}}-i f_{1{\rm T}}^\perp}{2}
\\
\frac{|{\bf k}_\perp| e^{i\varphi}}{M}\, \frac{g^\perp_{1{\rm T}}-i f_{1{\rm T}}^\perp}{2} &
\frac{{\bf k}_\perp^2 e^{i2\varphi}}{2 M^2}   h_{1{\rm T}}^\perp & \frac{f_1-g_1}{2}
&  \frac{-|{\bf k}_\perp|e^{i\varphi}}{M}\, \frac{h_{1{\rm L}}^\perp+i h_{1}^\perp}{2} \\
h_1 & \frac{-|{\bf k}_\perp| e^{i\varphi}}{M}\,\frac{g^\perp_{1{\rm T}}+i f_{1{\rm T}}^\perp}{2}  &
\frac{-|{\bf k}_\perp|e^{-i\varphi}}{M}\, \frac{h_{1{\rm L}}^\perp-i h_{1}^\perp}{2} &  \frac{f_1+g_1}{2} \\
\end{array}
\!\!\!\!
\right)\!\!(x,{\bf k}_\perp^2),
\end{eqnarray}
where column $a=\Lambda^\prime\lambda^\prime$ and row $b=\Lambda\lambda$ with $a,b \in \{\Rightarrow\rightarrow,\Rightarrow\leftarrow,\Leftarrow\rightarrow,\Leftarrow\leftarrow\}$ are labeled by
the proton($\Rightarrow\atop \Leftarrow$) and struck quark($\rightarrow\atop \leftarrow$) light cone spin projections.
 Here, we use standard notation, where the sets $\{f_1, g_1, h_1\}$,
$\{g^\perp_{1{\rm T}}, f_{1{\rm T}}^\perp, h_{1{\rm L}}^\perp, h_{1}^\perp\}$, and $h_{1{\rm T}}^\perp$   are associated with twist-two, twist-three, and
twist-four PDFs, respectively. This spin density matrix  possesses a certain symmetry, which can be used for a classification scheme of quark models. To proceed, we represent the spin density matrix (\ref{tPhi}) as the overlap of $n$-parton LCWFs, written as convolution
\begin{eqnarray}
\label{tPhi-LCWF-0}
\mbox{\boldmath $\widetilde \Phi$}(x,{\bf k}_\perp) = \sum_n \mbox{\boldmath $\widetilde \Phi$}_{(n)}(x,{\bf k}_\perp)\,,
\quad
\mbox{\boldmath $\widetilde \Phi$}_{(n)}(x,{\bf k}_\perp) =
\mbox{\boldmath $\psi$}^\ast_{(n)}(X_i,{\bf k}_{\perp i},s_i) \stackrel{(n-1)}{\otimes}  \mbox{\boldmath $\psi$}_{(n)}(X_i,{\bf k}_{\perp i},s_i)
\end{eqnarray}
that includes the phase space integration of $(n-1)$ spectators, the sum over their spin projections, and a direct product of LCWF ``spinors'', describing the struck quark--proton spin correlation
$$
\mbox{\boldmath $\psi$}_{(n)}(X_i,{\bf k}_{\perp i},s_i)=
\left(
  \begin{array}{c}
    \psi^{\Rightarrow}_{\rightarrow,n} \\
    \psi^{\Rightarrow}_{\leftarrow,n} \\
    \psi^{\Leftarrow}_{\rightarrow,n} \\
    \psi^{\Leftarrow}_{\leftarrow,n} \\
  \end{array}
\right)(X_i,{\bf k}_{\perp i},s_i)\,.
$$

As ``spherical'' models we denote those for which the spin density matrix (\ref{tPhi}) possesses triply degenerated eigenvalues, which are given by $(f_1 + g_1 -  2h_1)/2 $ or $(f_1 + g_1 + 2h_1)/2 $. In such uPDF models three linear  constraints and one quadratic constraint must be fulfilled,
\begin{eqnarray}
\label{constraint-rank1-1}
&&\!\!\!\!\!\!\!\!g_{1\rm T}^\perp \pm  h_{1\rm L}^\perp =0\,,
\quad
f_{1\rm T}^\perp \mp h_1^\perp =0\,,
\quad
g_1\mp h_1 \mp \frac{{\bf k}_\perp^2}{2 M^2} h_{1{\rm T}}^\perp  =0\,,
\\
\label{constraint-rank1-2}
&&\!\!\!\!\!\!\!\!\left(h^\perp_{1 \rm L}\right)^2 +\left(h^\perp_{1} \right)^2 + 2 h_1 h^\perp_{1 {\rm T}} = 0\,,
\end{eqnarray}
where the upper or lower sign has to be taken consistently. If the triple eigenvalues are non-zero, the spin density matrix has rank-four (or rank-three if the singlet eigenvalue vanish) and one needs four (or three)  overlap contributions that arise from independent LCWF ``spinors''.  On the other hand, if the positivity bound $h_1=\pm (f_1 + g_1)/2$ for uPDFs  (analogous to the Soffer bound for PDFs) is saturated, the triple eigenvalues vanish and the spin density matrix has rank-one, i.e., it can be represented by the direct product of one effective LCWF ``spinor'' as
\begin{eqnarray}
\label{bfPhi-rank1}
\mbox{\boldmath $\widetilde \Phi$}^{\rm rank-1}(x,{\bf k}_\perp) = \mbox{\boldmath $\widetilde \Phi$}_{(1)}(x,{\bf k}_\perp) = \left(
\psi^{\Rightarrow}_{\rightarrow},  \psi^{\Rightarrow}_{\leftarrow},  \psi^{\Leftarrow}_{\rightarrow},  \psi^{\Leftarrow}_{\leftarrow}
\right)^\ast \otimes
\left(
  \begin{array}{c}
    \psi^{\Rightarrow}_{\rightarrow} \\
    \psi^{\Rightarrow}_{\leftarrow} \\
    \psi^{\Leftarrow}_{\rightarrow} \\
    \psi^{\Leftarrow}_{\leftarrow} \\
  \end{array}
\right)(X,{\bf k}_{\perp})\,.
\end{eqnarray}
The ``spherical'' models for which the upper sign holds true are realized in scalar diquark, axial-vector diquark of \cite{JakMulRod97}, covariant parton \cite{EfrSchTerZav09}, bag \cite{AvaEfrSchYua10}, chiral quark soliton \cite{LorPasVan11}, and three-quark LCWF \cite{PasCazBof08} models. Indeed, up to the choice of a scalar LCWF, all these uPDF models can be considered as equivalent, even if they might have different struck quark--proton couplings. More generally, we may represent such models for a given quark species as
\begin{eqnarray}
\label{spin-matrix-spherical}
\mbox{\boldmath $\widetilde \Phi$}^{q}(x,{\bf k}_\perp) \stackrel{\rm sph}{=} \frac{1}{2} {\bf 1\!\!\!1}_{4\times 4} f_1^{q}(x,{\bf k}_\perp)+ \left[\left(
\mbox{\boldmath $\psi$}^{\ast\ \rm sca}\otimes \mbox{\boldmath $\psi$}^{\rm sca}\right) -  \frac{1}{4} {\bf 1\!\!\!1}_{4\times 4}
{\rm Tr}\, (\mbox{\boldmath $\psi$}^{\ast\ \rm sca}\otimes \mbox{\boldmath $\psi$}^{\rm sca})\right](x,{\bf k}_\perp)\,,
\end{eqnarray}
where $\mbox{\boldmath $\psi$}^{\rm sca}$ is the LCWF ``spinor'' of a ``(pseudo)scalar'' diquark model and the unpolarized uPDF $f_1^q$ is simply given as the overlap of a ``scalar'' LCWF. Note that in the aforementioned models only one scalar LCWF appears and that SU(4) flavor-spin symmetry ties the  uPDFs of $u$ and $d$ quarks.

As ``axial-symmetric'' model we denote those for which the spin density matrix has two degenerated eingenvalues. These doubly degenerated eigenvalues can arise from one pair of roots, i.e., $p_i \pm \sqrt{\cdots_i}$ with $\sqrt{\cdots_i}=0$ for $i\in\{1,2\}$,  or different pairs, e.g.,  $p_1 \pm \sqrt{\cdots_1}= p_2 \pm \sqrt{\cdots_2}$.

In the former case the three linear relations (\ref{constraint-rank1-1}) are satisfied, however, the quadratic one (\ref{constraint-rank1-2}) does not hold true. Moreover, if the uPDF Soffer bound is saturated, the spin density matrices of these models have rank-two (or two zero modes). An example is the scalar diquark model containing a gauge link \cite{MeiMetGoe07,BacConRad08}, in which the naive $T$-odd functions satisfy $f_{1{\rm T}}^\perp=h_1^\perp$.

In the latter case of ``axial-symmetric'' models one forth order relation is fulfilled  which in the case of two zero-modes reduces to two quadratic equalities:
\begin{eqnarray}
\label{constraint-rank2-1a}
(f_1+g_1- 2 h_1) \left(f_1-g_1+\frac{{\bf k}^2_\perp}{M^2} h_{1 \rm T}^\perp\right)-
\frac{{\bf k}^2_\perp}{M^2}  \left[
\left(g_{1\rm T}^\perp + h_{1\rm L}^\perp\right)^2+ \left(f_{1\rm T}^\perp - h_1^\perp\right)^2
\right] =0\,,
\\
\label{constraint-rank2-1b}
(f_1+g_1 + 2 h_1) \left(f_1-g_1 -\frac{{\bf k}^2_\perp}{M^2} h_{1 \rm T}^\perp\right)-
\frac{{\bf k}^2_\perp}{M^2}  \left[
\left(g_{1\rm T}^\perp - h_{1\rm L}^\perp\right)^2 + \left(f_{1\rm T}^\perp + h_1^\perp\right)^2
\right] =0\,.
\end{eqnarray}
Such models are realized in axial-vector diquark models with $f_{1\rm T}^\perp=h_1^\perp=0$ where
only the transverse polarization
of the diquark is taken into account \cite{BacConRad08} or the so-called quark-target model \cite{MeiMetGoe07}.

In the case that the eigenvalues are not degenerated, however, one zero mode (rank-three) appears we still have one
quadratic model relation. An example of such a model is an axial-vector diquark model of Ref.~\cite{BacConRad08} where the diquark possess two transversal  and one longitudinal polarization, however, the polarization tensor differs from that of the ``spherical'' axial-vector diquark model \cite{JakMulRod97}.
Even if the time-like polarization is taken into account, i.e., the polarization tensor is $-g_{\mu \nu}$, one still has a rank-three model and the same quadratic constraint is satisfied,
\begin{eqnarray}
\label{constraint-rank3}
(f_1+g_1+2 h_1) \left(f_1-g_1-\frac{{\bf k}^2_\perp}{M^2} h_{1 \rm T}^\perp\right)-
\frac{{\bf k}^2_\perp}{M^2}
\left(g_{1\rm T}^\perp - h_{1\rm L}^\perp\right)^2 =0\,.
\end{eqnarray}

\subsection{Models for zero-skewness GPDs in impact space}

\noindent
For zero-skewness GPDs with leading twist-two an  classification scheme analogous that for uPDFs holds true in the impact space.
The spin density matrix is now given by
\begin{eqnarray}
\label{bfF-b}
&&\widetilde {\bf F}(x,b) = \left(\!\!\!
\begin{array}{cccc}
\frac{H  +\widetilde H}{2}
&
i\, e^{i\varphi}\, \frac{\overline{E}^\prime_{\rm T}}{2}
&
- i\, e^{-i\varphi}\,\frac{E^\prime}{2}
&
\overline{H}_{\rm T}
\\
- i\, e^{-i\varphi}\, \frac{\overline{E}^\prime_{\rm T}}{2}
&
\frac{H -\widetilde H }{2}
&
e^{-i2\varphi} \widetilde H^{\prime\prime}_{\rm T}
&
-i\, e^{-i\varphi}\, \frac{E^\prime}{2}
\\
i\, e^{i\varphi}\, \frac{E^\prime}{2}
&
 e^{i2\varphi} \widetilde H^{\prime\prime}_{\rm T}
&
\frac{H  -\widetilde H }{2}
&
i\, e^{i\varphi}\, \frac{\overline{E}^\prime_{\rm T}}{2}
\\
\overline{H}_{\rm T}
&
i\, e^{i\varphi}\, \frac{E^\prime}{2}
&
-i\, e^{-i\varphi}\, \frac{\overline{E}^\prime_{\rm T}}{2}
&
\frac{H +\widetilde H }{2}  \\
\end{array}
\!\!\!
\right)\!(x,b)\,,
\end{eqnarray}
where $b=|{\bf b}|$, $\varphi$ denotes now the polar angle in the impact parameter space, and
\begin{eqnarray}
E^\prime = \frac{b}{M} \frac{\partial}{\partial b^2} E(x,b)\,,
\quad
\overline{E}^\prime_{\rm T} = \frac{b}{M} \frac{\partial}{\partial b^2}\left[ E_{\rm T} + 2\widetilde H_{\rm T} \right](x,b)\,,
\quad
\widetilde{H}_{\rm T}^{\prime\prime} = \frac{b^2}{M^2} \frac{\partial}{\partial b^2} \frac{\partial}{\partial b^2}\widetilde{H}_{\rm T}(x,b)
\,.
\nonumber
\end{eqnarray}
denote (dimensionless) derivatives of GPDs.
Comparing the uPDF spin density matrix (\ref{tPhi}) with the GPD one (\ref{bfF-b}), one reads off the following correspondences:
\begin{eqnarray}
\label{GPD02uPDF}
&&\!\!\!\!\!\!  H(x,b) \leftrightarrow f_1(x,{\bf k}_\perp)\,,\;\; \widetilde H(x,b) \leftrightarrow g_1(x,{\bf k}_\perp)\,,\;\;
 \overline{H}_{\rm T}(x,b) \leftrightarrow h_1(x,{\bf k}_\perp) \,,
\\
&&\!\!\!\!\!\!
 E^\prime(x,b) \leftrightarrow -\frac{|{\bf k}_\perp|}{M} f_{1\rm T}^\perp(x,{\bf k}_\perp) \,,
\;\;
\overline{E}^\prime_{\rm T}(x,b) \leftrightarrow -\frac{|{\bf k}_\perp|}{M} h_{1}^\perp(x,{\bf k}_\perp)\,,
\quad
\widetilde{H}_{\rm T}^{\prime\prime}(x,b)  \leftrightarrow \frac{{\bf k}^2_\perp}{2M^2}h_{1\rm T}^\perp(x,{\bf k}_\perp) \,.
\nonumber
\end{eqnarray}
Note the mismatch in the $T$-odd sector, where $T$-odd GPDs  $\eta \widetilde E$ and  ${\hat E}_{\rm T}$, somehow corresponding to $T$-even uPDFs $g_{1{\rm T}}^\perp$ and $h_{1{\rm L}}^\perp$, drop out in the forward case and $T$-odd  uPDFs  $f_{1\rm T}^\perp$ and $h_{1}^\perp$ correspond to $T$-even  GPDs $E$ and $\overline{E}_{\rm T}$, respectively.  We emphasize that these correspondences can certainly be used on a formal level to adopt the above uPDF classification scheme for GPDs, however, this does not mean that, e.g., $f_{1\rm T}^\perp$ (vanishing in any pure quark model) is related to GPD $E$ (which usually does not vanish). Generally, uPDFs and GPDs are independent projections on certain LCWF overlaps and formally only three sum rules should be fulfilled for twist-two related uPDFs:
\begin{eqnarray}
\label{H2q}
q(x,\mu^2) &\!\!\! = \!\!\!& \intk\, q(x,{\bf k}_{\perp})\,,
\end{eqnarray}
where PDFs  $q\in \{f_1,g_1,h_1\}$ are given at the boundary $\eta=0$ and $t=0$ of GPDs $F\in\{H,\widetilde H, H_{\rm T}\}$.

\subsection{GPD models}

\noindent
Because of ${\bf k}_\perp$-integration, the spin-density matrix of common GPDs will in general possess less symmetry than uPDFs or GPDs in impact space.
For a ``spherical'' model we expect that an analog of the two  ${\bf k}_\perp$-independent linear relations (\ref{constraint-rank1-1}) exist,
however, the third one, which is ${\bf k}_\perp$-dependent, might have an equivalent as integral relation, however, there might not exist quadratic relations
such as in (\ref{constraint-rank1-2},\ref{constraint-rank2-1a}--\ref{constraint-rank3}). In addition for a spherical model of rank-one the analog of a saturated uPDF Soffer bound exist. Indeed, for a ``spherical'' model of rank-one four GPD relations hold true, which allow expressing the chiral even GPDs by the chiral odd ones:
\begin{eqnarray}
\label{GPDmod-con1}
H(x,\eta,t) &\!\!\! \stackrel{{\rm sph}^3}{=} \!\!\! & \pm\left[
\overline{H}_{\rm T}(x,\eta,t) - \frac{t}{4 M^2}\widetilde{H}_{\rm T}(x,\eta,t)  -
\int_{-\infty}^t\!\frac{ dt^\prime}{4M^2}\, \widetilde{H}_{\rm T}(x,\eta,t^\prime)+\eta \widetilde E_{\rm T}(x,\eta,t)\right]\,,
\\
\label{GPDmod-con2}
E(x,\eta,t) &\!\!\! \stackrel{{\rm sph}^3}{=} \!\!\! & \pm\left[
E_{\rm T}(x,\eta,t)+ 2 \widetilde{H}_{\rm T}(x,\eta,t) -\eta \widetilde E_{\rm T}(x,\eta,t)
\right] \,,
\\
\label{GPDmod-con3}
\widetilde H(x,\eta,t) &\!\!\! \stackrel{{\rm sph}^3}{=} \!\!\! & \pm\left[
\overline{H}_{\rm T}(x,\eta,t) + \frac{t}{4 M^2}\widetilde{H}_{\rm T}(x,\eta,t)  +
\int_{-\infty}^t\!\frac{ dt^\prime}{4M^2}\, \widetilde{H}_{\rm T}(x,\eta,t^\prime)\right]\,,
\\
\label{GPDmod-con4}
\widetilde{E}(x,\eta,t)&\!\!\! \stackrel{{\rm sph}^3}{=} \!\!\! & \pm\left[
E_{\rm T}(x,\eta,t)-\frac{1}{\eta}\widetilde E_{\rm T}(x,\eta,t)\right]\,.
\end{eqnarray}
We add that in a scalar diquark model there exist seven linear relation among the eight twist-two GPDs, which are given as constraints for double distributions, see \cite{MueHwa11}.

\section{Conclusions and outlook}

\noindent
Based on symmetry properties and the number of zero-modes of the spin density matrix, we suggested a geometrical classification scheme for quark models that is applicable for leading power uPDFs and leading twist GPDs. In some special cases our classification scheme allows for a partonic interpretation that is tied to internal rotation symmetry. For instance, in the case of ``spherical'' symmetry the spin-density matrix commutes with a unitary matrix that is composed of a Melosh transform,  applied on the struck quark, and an arbitrary rotation of struck quark and proton spins. This invariance implies the relations (\ref{constraint-rank1-1},\ref{constraint-rank1-2}). Moreover, in the new spin basis the off-diagonal
spin components $\psi^{\Rightarrow}_{\leftarrow} = \psi^{\Leftarrow}_{\rightarrow}=0$ vanish in a scalar diquark model and so the new LCWFs are
invariant under rotation, for a detailed discussion and interpretation see  \cite{LorPas11}. However,  the saturation of the uPDF Soffer bound and the quadratic relation (\ref{constraint-rank3}), valid in the axial-vector model \cite{BacConRad08}, arise from the limitation of taking independent LCWF ``spinors'' rather than from rotation symmetry.

Although the idea of a classification scheme for quark models is trivial, the scheme itself might be useful.
For instance, the rather non-trivial statement that any ``spherical'' uPDF model, e.g., the three-quark LCWF model \cite{PasCazBof08}, can be obtained from an axial-vector diquark model \cite{JakMulRod97}. Thereby, its spin density matrix (\ref{spin-matrix-spherical}) can be represented by a scalar diquark model and an additional unpolarized uPDF. An analog construction with what we call minimal axial-vector diquark--quark coupling yields equivalent uPDF models and a consistent GPD model in which $H$ and $E$ GPDs are tied to each other. In such a model
the established ``pomeron'' behavior of sea quark GPD $H^{\rm sea}$ appears also in GPD $E^{\rm sea}$, where polynomiality is completed. This latter GPD could be  accessed in a single transverse proton spin asymmetry, measured in the hard exclusive electroproduction of photons and so such measurements provide insight into the quark orbital angular momentum carried by sea quarks.  Finally, we emphasize that LCWF models, belonging to  a certain uPDF class, may yield leading twist-two GPDs that belong to another class. Moreover, we expect that non-leading power or twist quantities evaluated in a ``spherical'' three-quark and two-quark LCWF model are becoming nonequivalent.


\end{document}